\documentstyle[sprocl]{article}

\def\Journal#1#2#3#4{{#1} {\bf #2}, #3 (#4)}

\def\NPB{{\em Nucl. Phys.} B}
\def\NPA{{\em Nucl. Phys.} A}
\def\PLB{{\em Phys. Lett.}  B}
\def\PRL{\em Phys. Rev. Lett.}

\def\PRC{{\em Phys. Rev.} C}

\def\be{\begin{equation}}
\def\ee{\end{equation}}
\def\bea{\begin{eqnarray}}
\def\eea{\end{eqnarray}}

\begin{document}

\title{Perturbative Pions and the Effective Range Expansion}

\author{Thomas D. Cohen}

\address{Department of Physics, University of Maryland,\\ 
College Park, MD 20742\\
E-mail cohen@physics.umd.edu}


\maketitle
\abstracts{In this talk the $Q$ counting scheme to implement effective
field theory is discussed.  It is pointed out that there are two small
mass scales in the problem $m_\pi$ and $1/a$ with $1/a \ll m_\pi$.  It is
argued that while the expansion based on $1/a$ being small compared to
the underlying short distance scales works well, the chiral expansion
may not.  The coefficients of the effective range expansion are
sensitive to the chiral physics and are very poorly described in $Q$
counting at lowest nontrivial order.  A ``shape function'' is introduced
which again is sensitive to pionic physics and insensitive to fitting procedures.
It is also poorly described  in $Q$ counting.}

\section{Introduction}

The $Q$ counting scheme introduced just over one year ago by Kaplan, Savage and
Wise (KSW) \cite{KSW} represents an important advance in the development of effective
field theory techniques for nuclear physics.  The approach is systematic,
it builds in approximate chiral symmetry and chiral power counting, it solves
Weinberg's \cite{Weinberg} ``unnatural scattering length problem'', and,
in principle, it provides {\it a priori} estimates of errors for observables since one works to fixed  order in $Q/\Lambda$.  
This last feature is extremely important since in Weinberg's initial formulation  \cite{Weinberg} it was unclear how to make such error estimates.
\footnote{Of course, the principal difference between $Q$ counting and the Weinberg scheme is that in
Weinberg's approach the potential is iterated to all orders whereas in $Q$ counting only the leading term is iterated to all orders; all other operators are treated perturbatively.  If $Q$ counting is valid, a scheme like  Weinberg's if it can be systemat

ically implemented will, at worst, add
uncontrolled higher order contributions which do not spoil the systematic error
 estimates.  Thus even if one is using Weinberg's approach, if one wants to make
simple error estimates one may resort to $Q$ counting for the error estimate, provided the system is in  a regime where it is valid.  This view of error 
estimation in the Weinberg scheme is featured prominently in the discussions
at this workshop.}   To a considerable
practical extent the  major advantage to using EFT technology as opposed to 
unsystematic models is the ability to specify the accuracy of one's 
predictions.
Thus, in this talk, I will focus entirely on $Q$ counting and not on the
many beautifully accurate calculations based on Weinberg's approach implemented with a finite cutoff \cite{cutcalcs}.

The scheme introduced by KSW may be divided up into two parts.  The first is
$Q$ counting and the second is a set of technical tricks to implement the $Q$
counting.  These technical tricks are rather unusual; they are based on the
so-called PDS scheme for doing subtractions in dimensional regularization.
Apart from the very peculiar prescriptions required (subtracting the poles as $d=3$!) the formalism is not completely transparent in terms of the physics.  Of course, provided the scheme is 
consistent and we are in the regime for which $Q$ counting is valid,  one should get the same results for any scheme which implements the $Q$ counting.  Two
other approaches to $Q$ counting have been tried---the OS scheme with
dimensional regularization of Mehan and Stewart\cite{MS} and a cutoff scheme
in configuration space \cite{CH1}.  All the schemes  give the same results 
at fixed order
in $Q$ counting.  Thus, ultimately the physics turns on whether or not the $Q$ counting
is working.  

The $Q$ counting scheme is straightforward:
\begin{eqnarray}
Q \, & \sim  \, 1/a \nonumber \\
Q \, & \sim  \, m_\pi \nonumber \\ 
Q \, & \sim  \, p \label{Qcount} 
\end{eqnarray}
where $a$ is the scattering length--either singlet or triplet, and p represents
external momenta.  For partisans of the OS scheme,  you can just as well use
$\gamma =\sqrt{M B}$ where $B$ is the magnitude of the binding energy, in place of $1/a$.  All other scales are assumed to be heavy and will collectively be
denoted as $\Lambda$.  The expansion implied by $Q$ counting is in $Q/\Lambda$.
Now the key point of $Q$ counting is that since $p$, $m_\pi$ and $1/a$ are
all of the same order, at any order in $Q$ counting we have $ 1/(p a)$ and
 $p/m_\pi$ to all orders.   
One important feature of $Q$ counting is that while one needs to iterate the
lowest order contact term to all orders to get a consistent result, all higher
order contributions, including those from the pion can be treated
 perturbatively.

 $Q$ counting has been used to calculate a number of observables \cite{KSW,MS,Qpred} and 
generally seems to have real predictive power.  At first sight this would appear
to rule out the possibility that $m_\pi \sim \Lambda$.  However, most 
``vanilla'' observables principally test the $1/(a \Lambda)$ part of the  
theory.  Clearly, it is important to identify observables which are principally
sensitive to the $m_\pi/\Lambda$ parts of the theory and to rigorously test 
the chiral expansion.  Recall that only the chiral part of the $Q$ counting
is really understood in terms of QCD.  In $Q$ counting, the small value of 
$1/a$  is treated as an essential  fact of life that we cannot ignore.  
>From the QCD level, however, this fact of life is seen 
as essentially an accident. In contrast, the chiral physics is understood directly in terms of the small 
quark masses in the QCD Lagrangian along with spontaneous symmetry breaking.  The central theme of this talk is that the  
effective range expansion---namely the expansion of $p \cot (\delta )$ as 
a power series in 
energy---is 
a good place to test whether the chiral part of $Q$ counting is under control.

 The effective
 range expansion (ERE) is a good place to look at pionic effects for a number of reasons.  The expansion
may be written:
\begin{equation} 
p \cot (\delta ) \, = \, -\frac{1}{a} \, +\, \frac{1}{2} r_e \, p^2 \, +
 \,v_2 p^4 \, + \, v_3 p^6 \, + \, v_4 p^8 + \,   \ldots
\end{equation}
Simple $Q$ counting shows that the scattering length term is order $Q^1$
while all other terms in the ERE are at least order $Q^2$.  When one includes
pions explicitly the same counting holds; all terms except the first are
${\cal O}(Q^2)$.  It is important to note the distinction between the $Q$ expansion
and the momentum expansion in the ERE.  They differ precisely because the $Q$
expansion has $k/m_\pi$ and $1/(k a)$ to all orders while in the
 momentum expansion they are multiplied out.  One immediately deduces that 
$v_n \sim Q^{-2 n + 2}$.   Moreover all of the $v_i$ coefficients in the expansion diverge
in the chiral limit of $m_\pi \rightarrow 0$.  Hence they are pion dominated
quantities and should provide a test of the chiral part of $Q$ counting.

\section{Scales in Nuclear Physics}

Before coming to the effective range expansion in this approach, it is
useful to look a bit more closely at the various scales underlying $Q$ counting.
Formally, there are two light scales intrinsic to the problem, $1/a$ and
$m_\pi$.  In $Q$ counting they are both formally of the same order namely
${\cal O}(Q)$.  But emprically for both the triplet and singlet channel,
\begin{equation}
m_\pi \gg 1/a
\end{equation}
with $m_\pi a \approx 4$ for the triplet channel and $m_\pi a \approx 16$
for the singlet channel.
This raises the following logical possibility;
\begin{eqnarray}
m_\pi \, & \sim & \, \Lambda \nonumber\\
1/a \, & \ll & \, \Lambda 
\label{cond}\end{eqnarray}
{\it i.e.}, that there is no scale separation between pionic scales 
and the ``short
distance scales'' but there is a good scale separation between them and $1/a$.

Of course, an immediate prejudice is that the first relation 
in eqs.~(\ref{cond})
must be wrong;  chiral scales are intrinsically long compared to typical
hadronic scales.   However, what is relevant here for $\Lambda$ is not hadronic
scales, but {\it nuclear} scales.  The relationship of nuclear scales to QCD is quite
obscure, but it is certainly true that typical nuclear mass scales are much lower
than typical hadronic scales.  If conditions in eqs.~(\ref{cond}) turn out
be true one would expect that the parts of the theory which depend on
$ 1/(a \Lambda)$ will work quite well, while the parts which depend on
 $m_\pi/\Lambda$ will converge slowly or not at all.

Note, if it turns out that $ m_\pi \sim \Lambda $, there is nothing in
principle wrong with the $Q$ counting formalism and PDS.  It would simply
not be useful for real world situations.  Of course, one could play God
and consider a world in which the pion is much lighter than in nature
and then one would have real predictive power.  In principle, if lattice
technology improves, one could calculate properties in such an artificial
world from first principles and  could use the $Q$ counting technology 
to make predictions for this world.

Before looking at explicit calculations, we should ask whether it is reasonable to suppose
that $m_\pi \sim \Lambda$.  Ultimately, this question comes down to whether
 $1/a$ and $m_\pi$ are the only light scales in nuclear physics.  The answer
appears to be ``no''.  Numerically:
$$\frac{1}{m_\pi} \approx 1.5 {\rm fm} \; \; a^s \approx -23 {\rm fm}  \; \;
 a^t \approx 6 {\rm fm} $$
where the superscript s (t) refers to the singlet (triplet) channel.  Compare
these with the effective ranges:
$$r_e^s \approx 2.7 {\rm fm} \; \; r_e^t \approx 1.6 {\rm fm} $$
It is apparent that $m_\pi r_e \sim 1$; if $r_e \sim 1/\Lambda$ there is
a serious potential problem.

  Of course, it is possible that the large numerical size of $r_e$ is itself a reflection of chiral physics.  For example, if 
$r_e \sim 1/m_\pi$ there would be no problem.  One can use $Q$ counting
itself to answer the question of how $r_e$ behaves.  At leading nonvanishing order it is given by  \cite{MS,CH1}
\begin{equation}
r_e \, = \, {\cal O}(\Lambda) \, + \,
 \frac{g_A^2 M}{4 \pi f_\pi^2 a^2 m_\pi^2} \, - \
\frac{g_A^2 M}{3 \pi f_\pi^2 a m_\pi} \, =\,{ \cal O}(Q^0)
\end{equation}
Note that although there {\it is} a chiral enhancement---the last two terms
diverge in the chiral limit---it is compensated for by factors of the scattering
length in the denominator.  Thus, one expects in the context of $Q$ counting
the effective range to be a  short distance scale.  In practice, however, it is
larger than $1/m_\pi$.  This suggests, but does not prove,  that the chiral
scale is not well separated from ``short distance''   scales.

There is another way to see that ``short distance'' scales may be comparable
 to the pion mass scale.  Consider the typical scales in so-called realistic
 N-N potential models, {\it i.e., those which fit the scattering data}.
If you  look at the non--one-pion-exchange part of the potential it is, 
in fact,
larger than the OPEP potential for a distance less than $\sim 1.5-2$ fm.  
Since $1/m_\pi$ is comparable to, or shorter than, this distance 
we again appear to have 
evidence that non-chiral supposedly short distance scales are comparable to
$1/M_\pi$.  One might argue that the central potential contains 
two-pion-exchange physics (suitably reparameterized) in the potential model.  However, if $Q$ counting is valid, that contribution is chirally suppressed.

This argument appears to be model dependent as it is based on ``typical'' potential models.
There is a model independent way to constrain the short 
distance physics.\cite{scal}  Consider {\it any} nonrelativistic potential,
including possible non-local potentials.  Write the potential as the sum of an OPEP potential and some short distance potential with the constraint that the short distance potential vanishes beyond some distance $R$:
\begin{eqnarray}
V(\vec{r},\vec{r'}) \, = \, V_{\rm OPEP}(\vec{r}) \, \delta(\vec{r} - \vec{r'})
 + V_{\rm short}(\vec{r},\vec{r'}) \nonumber \\ \nonumber \\ \; \; \;  {\rm with}  \; \; \; V_{\rm short}(\vec{r},\vec{r'}) = 0 \;\; \; {\rm for } \; \; \;  r, r' \, > \, R.
\end{eqnarray}
Now suppose that this potential is inserted to a Schr\"odinger equation 
and used
to solve for singlet phase shifts.  A remarkable theorem can then 
be proved, namely that if the
short distance potential fits the scattering length and effective range there
is a minimum value for $R$.  For real world values one can deduce the 
$R > $1.1 fm  \cite{scal}.  Moreover the derivation of this bound shows that it
is unsaturatable so one expects $R$ to be significantly  
more than than 1.1 fm.  From this one deduces that  substantial 
contributions to the scattering 
come from ``short distance'' contributions which come from separations  of
greater than 1.1 fm.  Recalling
that $1/m_\pi \approx$ 1.4 fm, we see immediately that there is no significant
scale separation between $m_\pi$ and the scales fixing the overall range of
the nonpionic part of the potential.

While the preceeding arguments do not decisively prove that $\Lambda \sim m_\pi$
they certainly show that it is not implausible.

\section{$Q$ Counting and Cutoffs}
Now the problem comes down to computing $p \cot (\delta )$.  This can be done
in PDS as in ref.~1. 
For the present purpose it is instructive to
consider the cutoff calculation and we take our discussion 
from ref.~5. 
The essential physical idea in this approach is to implement the separation 
of long distance physics from short distance physics directly in configuration 
space.  A radius, $R$, is introduced as a matching point between long and short distance
effects; renormalization group invariance requires that physical quantities 
must  be independent of $R$.  It is important, however, that $R$ be chosen
large enough so that essentially all of the effects of the short distance
physics is contained within $R$. The potential is divided into the sum of two pieces,
a short distance part which vanishes for $r>R$ and a long distance part which vanishes for 
$r<R$. 
At $R$, the information about short 
distance effects is entirely contained in the energy dependence of the 
logarithmic derivative (with respect to position) of the wave function at $R$.
Thus, provided we can  parameterize this information systematically, we can formulate
the problem in a way which is insensitive to the details of the short distance
part of the potential.  This insensitivity to the details of the short distance physics is at the core of why effective field theory works.
For $r>R$, the Schr\"odinger equation is solved subject to the boundary 
conditions at $R$. 
For s wave scattering, the wave function at $R$ may be parameterized as 
$A \sin (k r + \delta_0)$; the energy dependence of the logarithmic derivative
is independent of $A$ and can be expressed in terms of an expansion similar to 
an effective range expansion:
\begin{equation} 
p \cot (\delta_0) \, = \, -1/a_{\rm short} \, 
+ \, 1/2 \,r_e^0\,p^2 \, + \, v_2^0 \,  p^4 +
\, v_3^0 \, p^6 + \,v_4^0 \, p^8 + \ldots \label{effrange0}
\end{equation}
Power counting in $Q$ for s wave scattering can be implemented straightforwardly.
All of the coefficients in the preceding expansion are assumed to be order $Q^0$
except the first term ($-1/a_{\rm short}$ ) which will be taken to be order $Q^1$
to reflect the unnaturally large scale  of the scattering length.  Power counting
for the long range part of the potential simply follows Weinberg's analysis
 \cite{Weinberg}, with the proviso that the potentials are only used for $r >R$.
At order $Q^2$ in $p \cot (\delta )$, only the simple one-pion-exchange contribution
to the $V_{\rm long}$ contributes.  The power counting also justifies an iterative
solution of the Schr\"odinger equation for $r>R$ along the lines of a conventional 
Born series.  It differs from the usual Born series in that the boundary conditions
at $R$ are implemented.  Finally, $Q$ counting is used in expanding out the final
expression for $k \cot (\delta )$.
  
Carrying out this program gives the following expression for $k \cot (\delta )$ 
at order $Q^2$ for the ${}^1S_0$ channel 
\begin{eqnarray}
p \cot (\delta ) & = &-\frac{1}{a_0} \, + \, m_\pi^2 \, \left[d + \,
 \frac{g_A^2  M}{16 \pi f_\pi^2} \, \left (\gamma + \ln (m_\pi R) \right) \right ]
 \, \nonumber \\ \nonumber \\
& + &\, \frac{1}{2} \, r_e^0 \,p^2 - \,  \, \frac{g_A^2  M}{64 \pi a_0^2 f_\pi^2} \,
\left( \frac{m_\pi^2}{p^2} \right )
\ln \left
(1 + \frac{4 p^2}{m_\pi^2} \right ) \nonumber \\ \nonumber \\
 & + &  \, \, \frac{g_A^2 m_\pi M}{16 \pi a_0 f_\pi^2} \, 
 \left( \frac{m_\pi}{p} \right )\,
 \tan^{-1} \left ( \frac{2 p}{m_\pi} \right ) \,
 + \,  \frac{g_A^2 m_\pi^2 M}{64 \pi f_\pi^2} \, 
 \ln \left (1 + \frac{4 p^2}{m_\pi^2} \right ) 
 \label{kcotd1}\end{eqnarray}
The convention used here has $f_\pi$ = 93 MeV.  
Apart from well-known parameters from pionic physics, there are three
parameters---$a_0$, $d$ and $r_e^0$, where   
 $1/a_{\rm short}$ from eq.~(\ref{effrange0}) is rewritten as 
$1/a_0 + d m_{\pi}^2$
 with $1/a_0 \sim Q$, and $d m_\pi^2 \sim Q^2$.
 These parameters fix the energy dependence of the
wave function at the matching scale $R$; renormalization group invariance
 requires $d$ to depend on R logarithmically.
This form  is precisely equivalent to the calculation in PDS, provided that the
 following identifications are made between the coefficients used above and
those used in PDS with the notation of ref.~1. 

 $$\frac{4 \pi}{M} \, \frac{1}{-\mu + 1/a_0} \, = \, C_0 $$

$$\frac{1}{2} \, r_e^0 \,  =  \, \frac{C_2 M}{4 \pi } \, \left ( \mu^2 \, - \, 
 \frac{2 \mu}{a_0} \, + \, \frac{1}{a_0^2} \right ) $$
 
\begin{eqnarray}
& m_\pi^2 &\, \left[d \, + \, 
 \frac{g_A^2  M}{16 \pi f_\pi^2} \, \left (\gamma + \ln (m_\pi R) \right) \right] \, =  \nonumber \\ \nonumber \\
 \,
 & \frac{g_A^2 \, M}{16 \pi \, f_\pi^2} &\, \left ( m_\pi^2 \,   \ln
\left (\frac{m_\pi}{\mu} \right ) \, - \,
 m_\pi^2 \,  + \, \frac{1}{a_0^2} \, - \, 2 \frac{\mu}{a_0}  \, + \mu^2 \right )
\nonumber \\ \nonumber\\
&  +  &\, \frac{D_2 M}{4 \pi} \left ( m_\pi^2 \mu^2 \, -  \, 
\frac{2 m_\pi^2 \mu}{a_0} \, + \, \frac{m_\pi^2}{a_0^2} \right)  
\label{equiv}\end{eqnarray}

Formally, this is encouraging in the sense that it explicitly demonstrates the 
scheme independence of physical quantities.  At the same time, there is an
important hint of trouble which may lie ahead.  In doing the derivation
the matching scale, $R$ was taken to scale as $Q^0$ and the quantity $m_\pi R$
as order $Q^1$.  To obtain the final expression only the leading term in $m_\pi R$ is kept.  If $R \sim 1/m_\pi$ this is clearly problematic, and from the
previous discussion about potentials, we see that $m_\pi R \sim 1$.

It should also be stressed that the quantity $p \cot (\delta )$ is an extremely
useful observable to work with in $Q$ counting.  Unlike the amplitude itself,
there are no poles near $p=0$;  thus the issues of reorganizing the expansion
as in OS  \cite{MS} do not come up.  Moreover, the expression is valid near $p=0$
(assuming that $Q$ counting holds) so it should be useful for ultra low energy
scattering.  

\section{Low Energy Theorems}

One difficulty with eq.~(\ref{kcotd1}) is that it is given in terms of $a_0$
which is the scattering length for the short distance potential only; as such it is not an observable.  However, one can express everything in terms of physical observables and in doing so develop ``low energy theorems''  \cite{CH2}.  The trick is  to rel
ate $a_0$ to the
physical scattering length as follows: 
\begin{eqnarray}
 - \frac{1}{a} \, & = &\, - \frac{1}{a_0} + \, m_\pi^2 \, \left[ d \, + \, 
 \frac{g_A^2  M}{16 \pi f_\pi^2} \, \left ( \gamma + \ln (m_\pi R) \right) \right ]
 \, + \, \frac{g_A^2  M}{16 \pi f_\pi^2} \, \left( 
 \frac{2 m_\pi}{a_0} \, - \, \frac{1}{a_0^2} \right ) \nonumber \\ \nonumber \\ \, &  = & \,  
 - \,\frac{1}{a_0} + {\cal O}(Q^2/\Lambda)
 \label{scat}
 \end{eqnarray}
Therefore in all of the  ${\cal O}(Q^2)$ terms in   eq.~(\ref{kcotd1}) one can
 replace $a_0$ by the physical $a$;  the error in doing this is ${\cal O}(Q^3)$
which is one order beyond the order at which I am working.
One gets the following expression orginally derived in ref.~8. 
\begin{eqnarray}
p \cot (\delta ) & = &-\frac{1}{a_0} \, + \, m_\pi^2 \, \left[d + \,
 \frac{g_A^2  M}{16 \pi f_\pi^2} \, \left (\gamma + \ln (m_\pi R) \right) \right ]
 \, \nonumber \\ \nonumber \\
& + &\, \frac{1}{2} \, r_e^0 \,p^2 - \,  \, \frac{g_A^2  M}{64 \pi a^2 f_\pi^2} \,
\left( \frac{m_\pi^2}{p^2} \right )
\ln \left
(1 + \frac{4 p^2}{m_\pi^2} \right ) \nonumber \\ \nonumber \\
 & + &  \, \, \frac{g_A^2 m_\pi M}{16 \pi a f_\pi^2} \, 
 \left( \frac{m_\pi}{p} \right )\,
 \tan^{-1} \left ( \frac{2 p}{m_\pi} \right ) \,
 + \,  \frac{g_A^2 m_\pi^2 M}{64 \pi f_\pi^2} \, 
 \ln \left (1 + \frac{4 p^2}{m_\pi^2} \right ) 
 \label{kcotd2}\end{eqnarray}

One can expand this as a Taylor series in $p$ to obtain ERE coefficients.
They are given by:
\begin{eqnarray}
v_2 \, &  = & \, \frac{g_A^2 M}{16 \pi f_\pi^2} \, \left ( \, -\frac{16}{3 a^2
\,
m_\pi^4}\,  + \, \frac{32}{5 a \,m_\pi^3} \, - \,\frac{2}{m_\pi^2} \right
)\nonumber \\ \nonumber \\
v_3 \, & = & \, \frac{g_A^2 M}{16 \pi  f_\pi^2} \, \left ( \, \frac{16}{ a^2 \,
m_\pi^6}\,  - \, \frac{128}{7 a \, m_\pi^5} \, + \,\frac{16}{3 m_\pi^4} \right )
\nonumber \\ \nonumber \\
v_4 \, & = & \, \frac{g_A^2 M}{16 \pi  f_\pi^2} \, \left
( \, -\frac{256}{5 a^2 \, 
m_\pi^8}\,  + \, \frac{512}{9 a \,m_\pi^7} \, - \, \frac{16}{ m_\pi^6} \right )
 \nonumber \\  \nonumber \\
& \ldots &
\label{vi}
\end{eqnarray}
The preceding low energy theorems are valid to leading nontrivial order in $Q$ counting; 
corrections are of {\it relative} order $Q/\Lambda$.

Several features of these expressions are notable.  The first is that they are
true predictions, independent of choices made in fitting, to this order in $Q$
 counting.  As such they are low energy theorems.  One consequence of this
is that different schemes ({\it eg.} fitting out the pole as in OS rather 
than using
$1/a$) will only give corrections at relative order $Q/\Lambda$. 
In this sense these predictions can be considered ``low energy theorems '' which 
become exact in the limit $(m_\pi,1/a)/\Lambda \rightarrow 0$. The 
second significant fact is that all terms for all the expressions for the $v_i$
coefficients diverge in the chiral limit of $m_\pi \rightarrow 0$.  This implies that these
quantities are dominated by pionic effects and hence are a good place to test
 whether the pionic parts of $Q$ counting are working.

If the pionic parts of $Q$ counting were under control one would expect
that these predictions would work well.  In practice, however, they work quite poorly.
This can be seen in Table~(\ref{LET}) where the prediction from the low energy theorems are compared with coefficients extracted from the Nijmegen partial wave analysis (PWA).  The prediction from the low energy theorems  are typically off by a factor of 

5 or so.   This suggests that pionic
parts of the $Q$ counting are failing rather badly.  
\begin{table}[t]
\caption{A comparison of the predicted effective range expansion 
coefficients, $v_i$,   for the ${}^1S_0$ 
and ${}^3S_1$ 
channels with coefficients extracted
from the Nijmegen partial wave analysis.\label{LET}}  
\vspace{0.2cm}
\begin{center}
\footnotesize
\begin{tabular}{|c| c | c | c |}
\hline
 &  $v_2$ (${\rm fm}^{3})$ & $v_3$ (${\rm fm}^{5})$ & $v_4$ (${\rm
fm}^{7})$\\
\hline \hline
$\delta$ (${}^1S_0$  channel)& &  & \\
 \hline \hline & & &\\
low energy theorem &  -3.3 & 17.8 & -108. \\ \hline & & & \\
partial wave analysis & -.48 & 3.8 & -17. \\ 
\hline \hline
$\delta$ (${}^3S_1$  channel)& &  & \\
 \hline \hline & & & \\
low energy theorem &  -.95 & 4.6  & -25.  \\  \hline & & & \\
partial wave analysis & .04 & .67 &  -4.0\\ 
\hline 
 \end{tabular} 

\end{center}
\end{table}

One possible difficulty with the comparison of the $v_i$ coefficients from
the low energy theorems with the experimental data is that there is no
experimental data.  The coefficients extracted from the Nijmegen PWA were
based on a fit to the smoothed ``best fit''.  In principle one should do
this fit including an error analysis based on the uncertainties.  This cannot
be done from the published data of the Nijmegen group as they did not publish information about correlated errors.  Thus, one might wonder whether it is meaningful
to extract high derivatives which are presumably rather sensitive to errors.
A simple error estimate in ref.~4 
concludes that the errors are
likely to be too large to get any quantitative information about the $v$
coefficients.  From this one might
conclude the disagreement between the low energy theorems and the ``data''
in Table~(\ref{LET}) is due to an inability to extract the $v$ from the 
scattering data in a reliable way.  This is almost certainly not the case,
however. 

The Nijmegen group made several independent fits
 to the scattering data.  One was the PWA.  The others were various potential models which were fit
directly to the data ({\it i.e.} not to the PWA).  These fits had a $\chi^2$
per degree of freedom of essentially unity.  Thus they can be regarded as
 independent fits to the data \cite{Nij2}.  As the potential models have different forms
 from each other, they clearly have different systematic errors.  Moreover in doing
 the least squares fit different models make different compromises in fitting 
individual data points so that they tend to explore the statistical errors.
Thus, one might expect that the spread in the coefficients extracted in the
 different fits gives a reasonable feel for the  scale of the uncertainty.  
Table~(\ref{pot}) shows the $v_i$ coefficients for these fits for the triplet channel \cite{Stoksa},  and it is manifestly
clear that the spread in the effective range parameters as extracted from the
 three is vastly smaller than the difference with the predictions from the low energy theorems.

\begin{table}[t]
\caption{A comparison of the effective range expansion 
coefficients, $v_i$,   for the ${}^3S_1$ 
and ${}^3S_1$ 
channel predicted from the low energy theorem with coefficients extracted
from the  partial wave analysis and with three potential models---Nijmegen I, Nijmegen II and Reid 93---which were 
 fit directly to the scattering data. \label{pot}}
\vspace{0.2cm}
\begin{center}
\begin{tabular}{|c| c | c | c |}
\hline
$\delta$ (${}^3S_1$  channel)&$v_2$ (${\rm fm}^{3})$ & $v_3$ (${\rm fm}^{5})$ & $v_4$ (${\rm
fm}^{7})$\\ \hline \hline & & & \\
low energy theorem &  -.95 & 4.6  & -25.  \\  \hline & & & \\
partial wave analysis & .040 & .672 &  -3.96\\ 
Nijm I                & .046 & .675 & -3.97 \\
Nijm  II              & .045 & .673 & -3.95 \\
Reid93                & 0.033 & .671 & -3.90 \\
\hline 
 \end{tabular} 

\end{center}
\end{table}

\section{ Re-summing the Effective Range Expansion}

The effective range expansion parameters discussed in the previous section
provide a dramatic way to see that the pionic parts of $Q$ counting may
have serious problems with convergence.  However, there are a number of drawbacks
with looking at the $v_i$ coefficients.  As noted above, there are ambiguities
 in the extraction from the data and it is hard to get reliable error bars. Moreover, the effective range expansion itself
has a very limited radius of convergence.  Because of the pion cut one expects
the effective range expansion to converge only for $p < m_\pi /2$.

Of course, all of the low energy theorems for the $v_i$ coefficients are 
contained in eq.~(\ref{kcotd2}).  We can study this directly without expanding
as a function of $k$.  In effect, this amounts to re-summing the effective
range  expansion and using $p \cot (\delta )$ as our fundamental quantity.  There are two obvious advantages to doing this:  First, one can
avoid the problem of extracting the $v_i$ coefficients from noisy data and instead we can  compare directly with the partial wave analysis (which includes error estimates).  Second, one is no longer restricted to $p < m_\pi /2$ since the
re-summed expression is valid over the entire domain of $Q$ counting.  Unfortunately, $p \,cot (\delta )$ does not isolate the pionic contributions 
from the rest,
and at low $p$ it is dominated by $1/a$ physics and the fitting procedure
which gives the effective range.  There is a clean way to finesse this problem,
however.  Rather than study $p \cot (\delta)$ directly, one can study the
following ``shape function''\cite{CH3}:
\begin{equation}
{\cal S}(p) \, = \, p \cot (\delta) - (-1/a + 1/2 r_e p^2)
\end{equation}
which is just the re-summed effective range expansion with the first two terms
subtracted off.  This has the advantage of removing completely the sensitivity
to $1/a$ and the fitting of the effective range.   The quantity is completely pion dominated since in a theory with pions
integrated out ${\cal S} (p)$ is ${\cal O}(Q^3)$, while in a theory with
explicit pions it is ${\cal O}(Q^2)$.

  In Table ~(\ref{S}) we see the low energy
theorem prediction for ${\cal S} (p)$ compared with values extracted from the
Nijmegen partial wave analysis for the triplet channel.  Note the disagreement
is quite pronounced.  Moreover, note that error estimates are given for the
results extracted from the scattering data.  It is manifestly clear that
the discrepancies are {\it not} due to uncertainties in the data.  Again this suggests that the pionic parts of $Q$ counting are not predictive at leading nontrivial order---at least not for
this observable.

\begin{table}[t]
\caption{A comparison of the shape function ${\cal S}(p^2) =
p \cot (\delta) + 1/a - 1/2 r_e p^2 $ for the ${}^3 S_1$ channel extracted
from the Nijmegen partial wave analysis with the prediction by the low energy theorem. \label{S}}
\vspace{0.2cm}
\begin{center}
\begin{tabular}{|c  |c | c|}
\hline
lab energy (MeV) &${\cal S}$ extracted (Mev)
& ${\cal S}$ low energy theorem (Mev)\\
\hline \hline
Deuteron Pole &  $-0.0017 \pm  0.0125$ & -0.743\\
1  &  $-0.00095 \pm 0.00721$ & -0.0258 \\
5  & $0.0428 \pm 0.0194$ & -0.535 \\
10 & $0.245 \pm 0.047$  &  -1.78 \\
25 &  $ 2.18 \pm 0.14$ & -7.54 \\
50 & $  11.03 \pm 0.24 $ & -20.10\\
\hline 
\end{tabular}
\end{center}
\end{table}

\section{Conclusions}

The development of $Q$ counting is an extremely important step in our
understanding of effective field theory for nuclear physics.  However,
as stressed in this talk $Q$ counting involves two small mass parameters, $1/a$
and $m_\pi$.  While there is every indication that the part of the theory based
on expanding in $1/(a \Lambda)$ is working well,  the chiral counting is far 
more problematic.  The expansion has no predictive power for the 
effective range  parameters and the shape function
at leading nontrivial order (NLO).  As both of these quantities are 
chirally  sensitive this failure suggests that the chiral expansion 
may not be well under control.

There are a number of possibilities.  The most optimistic one is simply that
the NLO calculation is not adequate, and if one works at higher order all
will be well.  The most pessimistic possibility is that that the chiral expansion is not
converging and that this failure is general.  Clearly, the way to resolve the
situation is to work at higher order.  It is important when doing so, however,
to focus on observables which are highly sensitive to the pion physics.

\section*{Acknowledgments}
Most of the work discussed in this talk was done in collaboration
with James Hansen.  The research was funded by the U.S. Department of
Energy under grant no. DE-FG02-93ER-40762.

\end{document}